\documentclass[aps,prb,twocolumn,unsortedaddress,amsmath,amssymb]{revtex4-1}

\usepackage[bookmarks,bookmarksopen]{hyperref}
\usepackage{graphicx}
\usepackage{dcolumn}
\usepackage{bm}
\usepackage{color}
\def\ref#1{$^{#1}$}

\begin{document}

\preprint{APS}
\preprint{Submitted to J.\ Chem.\ Phys.}

\title{Notes on {\em ab initio} investigation of the CrH molecule and its interaction with He}

\author{Jacek K{\l}os}
\affiliation{Department of Chemistry and Biochemistry, University of Maryland, College Park, MD 20742-2021, USA}

\author{Micha{\l} Hapka and Grzegorz Cha{\l}asi{\'n}ski}
\affiliation{Faculty of Chemistry , University of Warsaw, ul. Pasteura 1, 02-093, Warsaw, Poland}

\date{\today}

\begin{abstract}
Potential and dipole moment curves for the CrH(X$^6\Sigma^+$) molecule were obtained at the internally-contracted multi-reference configuration interaction with single and double excitations and Davidson correction ({\em ic}-MRCISD+Q)  level using large basis set augmented with additional diffused functions and using Douglass-Kroll Hamiltonian for scalar relativistic effects. Also bound states, average positions and rotational constants calculated on the CrH(X$^6\Sigma^+$) potential are reported.
The He-CrH(X$^6\Sigma^+$) potential energy surface was calculated with the coupled cluster singles, doubles, and noniterative triples [RCCSD(T)] method. The global minimum was found for the collinear He$\cdots$Cr-H geometry with the well depth of 1143.84 cm$^{-1}$ at $R_e=4.15$ a$_0$. An insight in the character of the complex was gained by means of symmetry-adapted perturbation theory (SAPT) based on DFT description of the monomers. The presence of the so called ``exchange cavity'' was observed.
Finally, bound states of the He-CrH complex for $J$ = 0 are presented.

\end{abstract}

\maketitle

\section{Introduction}
The CrH molecule is a focus of interest in the astrophysics and cold molecule research. In the astronomy and astrophysics CrH molecule was identified in sunspots and is used to perform clasification of L-type brown dwarf stars into its subtypes.
Regarding the cold molecules, the CrH molecule along with MnH one was successfully buffer gas cooled recently by M. Stoll and others in  Gerard Meijer's group~\cite{stoll:2008}. The CrH molecule is paramagnetic and can have a lifetime of 0.12 seconds while trapped in buffer gas $^3$He at temperature of 0.65K~\cite{stoll:2008}.

\section{X$^6\Sigma^+$ and $^8\Sigma^+$ electronic states of the CrH diatomic}
\label{sec:crh} 
We focus on the ground electronic state of the CrH which is X$^6\Sigma^+$ and originates
from the $3d^54s^1$ Cr valence shell and $1s$ of H atom forming $4s\sigma^23d\sigma^13d\delta^23d\pi^2$ electronic configuration. In our {\em ab initio} approach to calculate potential and dipole moment
we used multi-configurational self-consistent field (MCSCF) to obtain reference orbitals for subsequent internally contracted multi-reference configuration interaction calculations including explicitly single and double excitations ({\em ic}-MRCISD). Davidson correction was applied to account for effects of higher excitations in an approximate manner. The Cr and H atoms were described by all-electron correlation consistent quadruple-zeta basis sets designed for Douglass-Kroll relativistic calculations (aug-cc-pVQZ-DK)~\cite{dunning:89,balabanov:2005}. A better description of the partial anionic character of the CrH dipole was achieved by adding $spdf$ diffuse functions placed on H atom with the following exponents: $s,0.00788$, $p,0.0283$, $d,0.063$ and $f,0.13$. 
 
\begin{figure}[h]
\includegraphics[width=8.0cm]{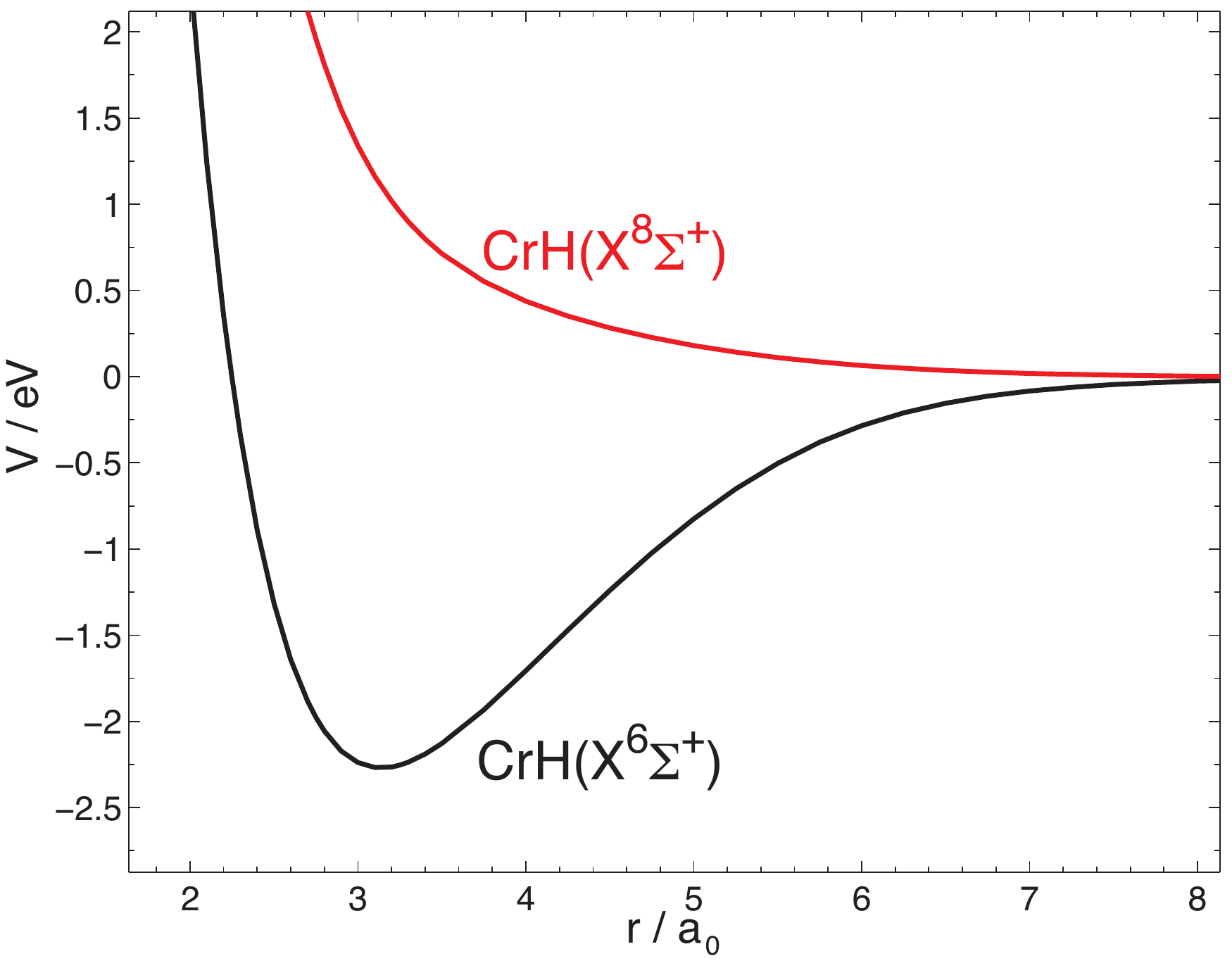}
\caption{{\em ic}-MRCISD+Q(Davidson) potential energy curves for the ground  X$^6\Sigma^+$ and spin-polarized $^8\Sigma^+$ CrH molecule. Note units are in eV.}
\label{crh_6s_8s}
\end{figure}

The reference wave function for the MCSCF calculations were obtained from the restricted Hartree-Fock calculations (RHF) for the high-spin case. The first step in the  MCSCF calculations was to perform state-averaged calculations for the X$^6\Sigma$ and $^8\Sigma$ states. This formed a starting point for subsequent single-state MCSCF calculations of the ground electronic states along the dissociation variable $r$, except for the results presented in Figure~\ref{crh_6s_8s} when we obtained both states. The active space in the MCSCF calculation was composed of 11 orbitals in symmetry $A_1$, 3 orbitals of $B_1$ and $B_2$ symmetry each, and 1 orbital of $A_2$ symmetry. We kept $3s$ and $3p$ orbitals correlated and always doubly occupied.   

In Figure~\ref{crh_6s_8s} we show the diatomic potential for the ground X$^6\Sigma^+$ state and for the spin-polarized octet $^8\Sigma^+$ as well. The high-spin state is practically repulsive in this plot in comparison to the X state. In our calculations the well depth, $D_e$, of the X$^6\Sigma^+$ state is 18485.21 cm$^{-1}$ (2.292 eV) at $r_e=3.111$ a$_0$. Our minimum is slightly deeper than one of 2.11 eV obstained by Dai and Balasubramanian~\cite{dai:1993}. The equilibrium distance agrees reasonably with the experimental one of 3.1275 a$_0$ measured by Bauschlicher et al in 2001~\cite{bauschlicher:2001}. Using the discrete variable representation (DVR) approach we calculated vibrational bound states for the CrH(X$^6\Sigma^+$) molecule which are presented in Table I along with average position $<r>$ and rotational constants  $B_v$. The zero-point energy of the CrH(X) is $D_0$=17626.88 cm$^{-1}$ which is 2.185 eV. This agrees reasonably well with the experimental $D_0^{exp}=1.93\pm0.07$ eV of Bauschlicher~\cite{bauschlicher:2001}.
 
\begin{figure}[h]
\includegraphics[width=8.0cm]{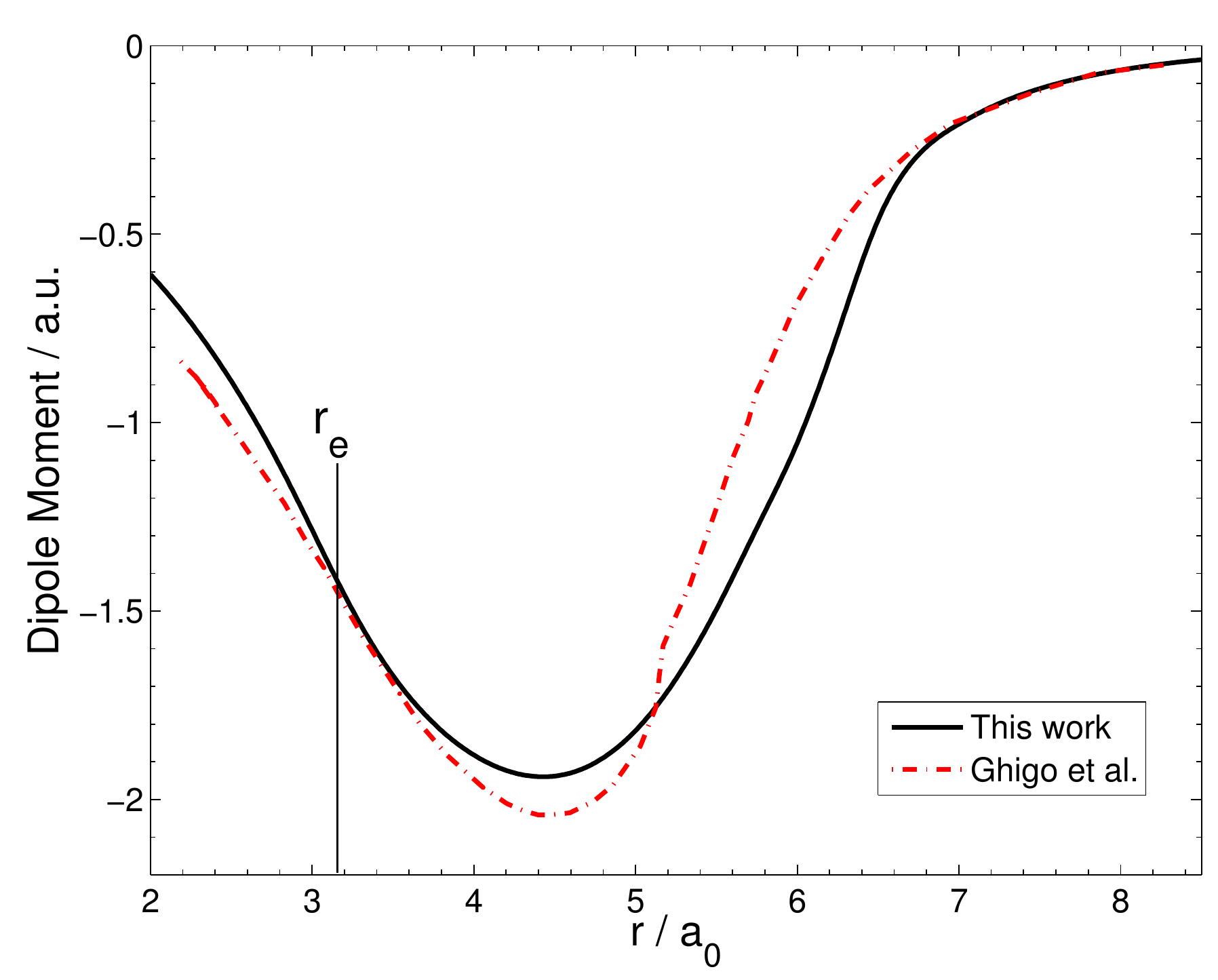}
\caption{Solid black line: {\em ic}-MRCISD dipole moment function of the ground  X$^6\Sigma^+$  state of the CrH molecule. Red dashed line: Values obtained from Ghigo {\em et al}~\cite{ghigo:2004}. Vertical line indicates equilibrium position $r_e$ of the CrH(X) molecule.}
\label{crh_6s_dipole}
\end{figure}

The dipole moment of the CrH(X) state is very large, 1.382 a.u. (3.512 D) at the equilibrium position. We calculated the dipole moment function using the expectation value of the dipole moment operator with {\em ic}-MRCISD density. Figure~\ref{crh_6s_dipole} shows the dipole moment curve compared to results of Ghigo {\it et al.}~\cite{ghigo:2004} who applied the multistate CASPT2 method based on reference wave functions from the state-averaged CASSCF with 16 molecular orbitals in the active space. We observe a good agreement between our {\em ic}-MRCISD and CASPT2 results for the dipole moment function. 

\begin{table} \small
\caption{Vibrational energy levels, average position and rotational constants for the non-rotating ground electronic state of the CrH molecule. Energies and rotational constants in cm$^{-1}$}
\begin{tabular}{cccc}
\hline\hline
$v$ & $E_v$ & $\left<r\right>$ / a$_0$ & $B_v$\\
\hline
   0 &  -17626.88 & 3.150 & 6.203\\
   1 &  -15972.38 & 3.235 & 6.014\\
   2 &  -14403.21 & 3.325 & 5.821\\
   3 &  -12909.22 & 3.419 & 5.628\\
   4 &  -11485.27 & 3.517 & 5.436\\
   5 &  -10127.77 & 3.620 & 5.246\\
   6 &   -8835.44 & 3.728 & 5.054\\
   7 &   -7607.84 & 3.843 & 4.859\\
   8 &   -6445.99 & 3.967 & 4.660\\
   9 &   -5352.01 & 4.102 & 4.451\\
  10 &   -4329.34 & 4.253 & 4.230\\
  11 &   -3383.31 & 4.426 & 3.990\\
  12 &   -2521.76 & 4.632 & 3.721\\
  13 &   -1758.37 & 4.901 & 3.392\\
  14 &   -1118.88 & 5.271 & 2.982\\
  15 &    -608.05 & 5.745 & 2.550\\
  16 &    -254.49 & 6.581 & 1.925\\
  17 &     -55.38 & 8.000 & 1.264\\
\hline
\end{tabular}
\end{table}

\section{RCCSD(T) Potential Energy Surface  for  the He-CrH(X$^6\Sigma^+$) Complex}
\label{sec:hecrh}
The potential energy surface for the He-CrH(X) complex was calculated at the coupled cluster singles, doubles, and noniterative triples, RCCSD(T), level of theory. The reference wave function for the RCCSD(T) calculations was obtained by first performing two-state averaged CASSCF calculations, saving pseudo-canonical orbitals for the ground state and then starting RHF calculations. We used aug-cc-pwCVQZ-DK basis set for Cr atom and aug-cc-pVQZ-DK basis sets for H and He. In RCCSD(T) calculations $1s$, $2s$, $2p_y$, $2p_z$ and $2p_x$ core orbitals of Cr were frozen.  

\begin{figure}
\includegraphics[width=8.0cm]{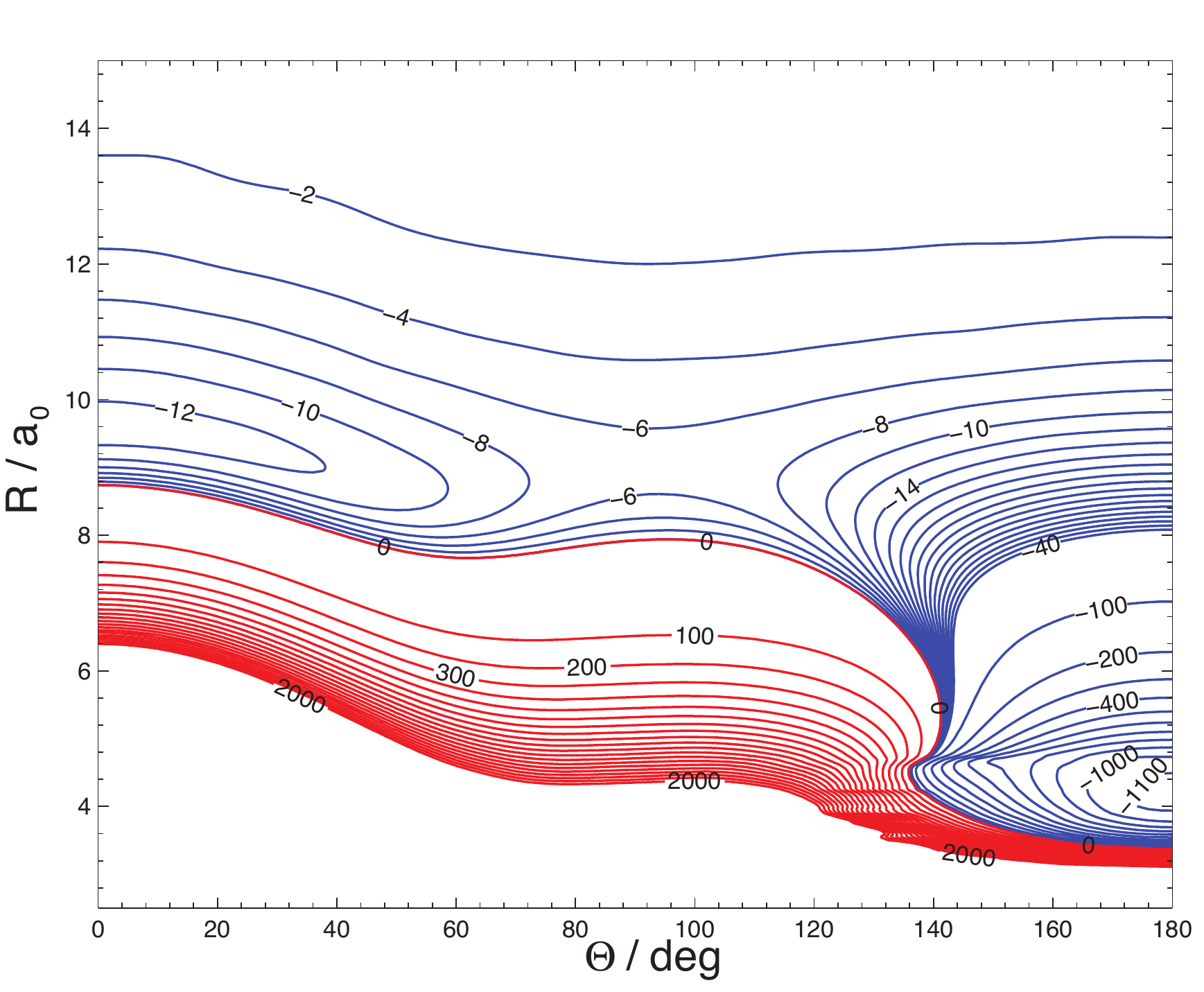}
\caption{The RCCSD(T) potential energy surface for the CrH(X$^6\Sigma^+$)-He van der Waals system. Contour labels are units of cm$^{-1}$ }
\label{crh_6s_he_pes}
\end{figure}

The contour plot of the He-CrH(X) potential is shown in Figure~\ref{crh_6s_he_pes}. The global minimum with a well depth of $D_e=1143.84$ cm$^{-1}$ is located at $R_e=4.15$ a$_0$ for $\theta_e=180^{\circ}$ which corresponds to the collinear He$\cdot\cdot\cdot$Cr-H geometry. There is an additional local minimum at He$\cdot\cdot\cdot$H-Cr linear geometry ($\theta=0^{\circ}$) located at $R=9.6$ a$_0$ with a well depth of 12.75 cm$^{-1}$. A T-shaped saddle point occurs at $R=9.05$ a$_0$ and $\theta=94^{\circ}$ with a barrier height of -6.6 cm$^{1}$. The T$_1$ diagnostic was on the order of 0.11 and D$_1$ diagnostic 0.44.
 
 \begin{figure}
\includegraphics[width=8.0cm]{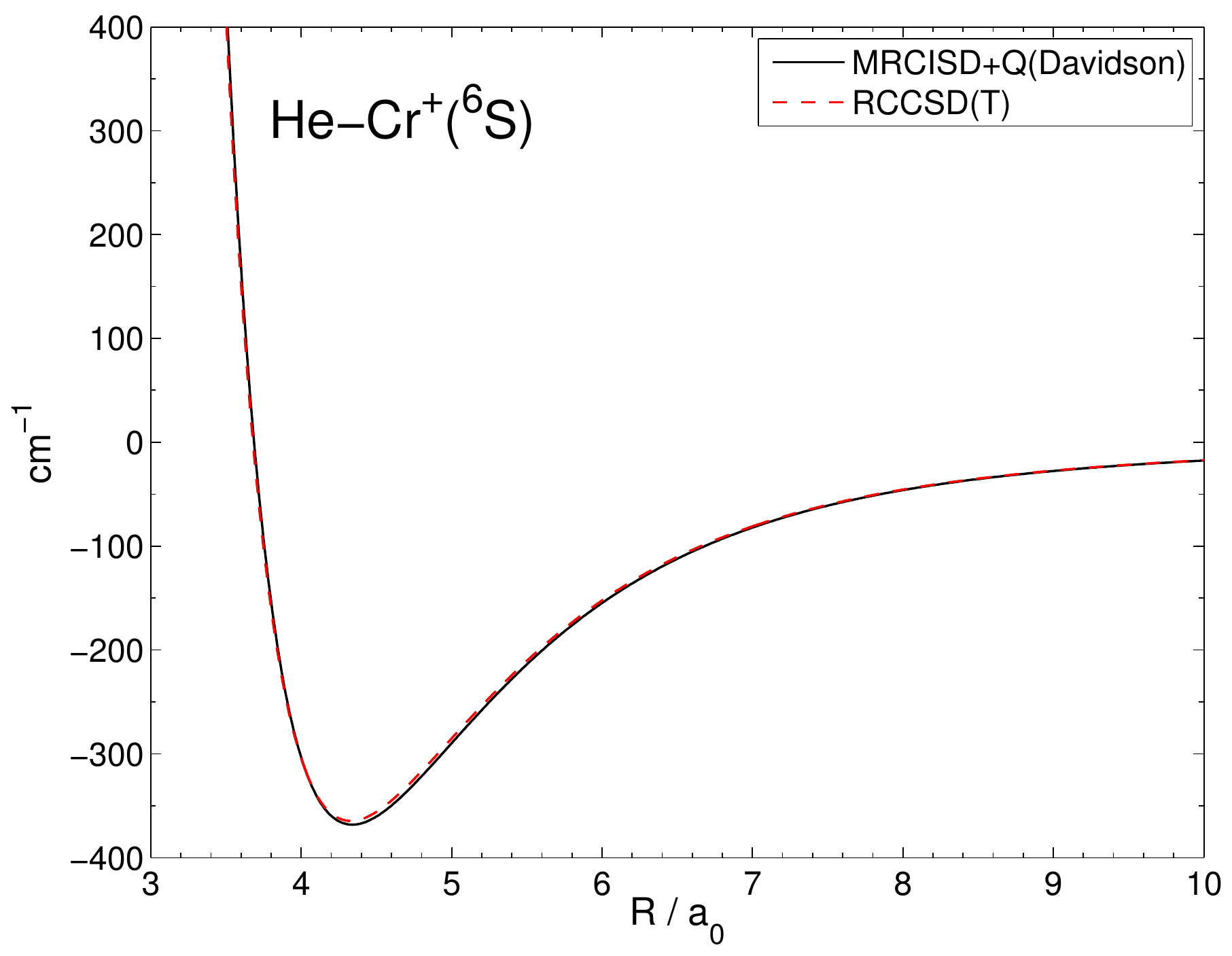}
\caption{The {\em ic}-MRCISD+Q(Davidson) (black solid) and RCCSD(T) (red dashed) potential energy curves  for the He-Cr$^+$($^6$S) van der Waals system. }
\label{hecr_plus}
\end{figure}

We compare the RCCSD(T) potential with the potential obtained from RHF+dispersion hybrid approach, combining the HF supermolecular energies with SAPT(UKS) dispersion contribution, where SAPT(UKS) stands for symmetry-adapted perturbation theory based on unrestriced Kohn-Sham description of the monomers. \cite{hapka2012}. The SAPT(UKS) dispersion energy has been calculated in the nonrelativistic framework with the PBE0 xc functional. \cite{Perdew:1996, Adamo:99} The aug-cc-pVQZ basis set with additional diffuse functions on the H atom has been chosen, as described in the previous section.

\begin{figure}
\includegraphics[width=8.0cm]{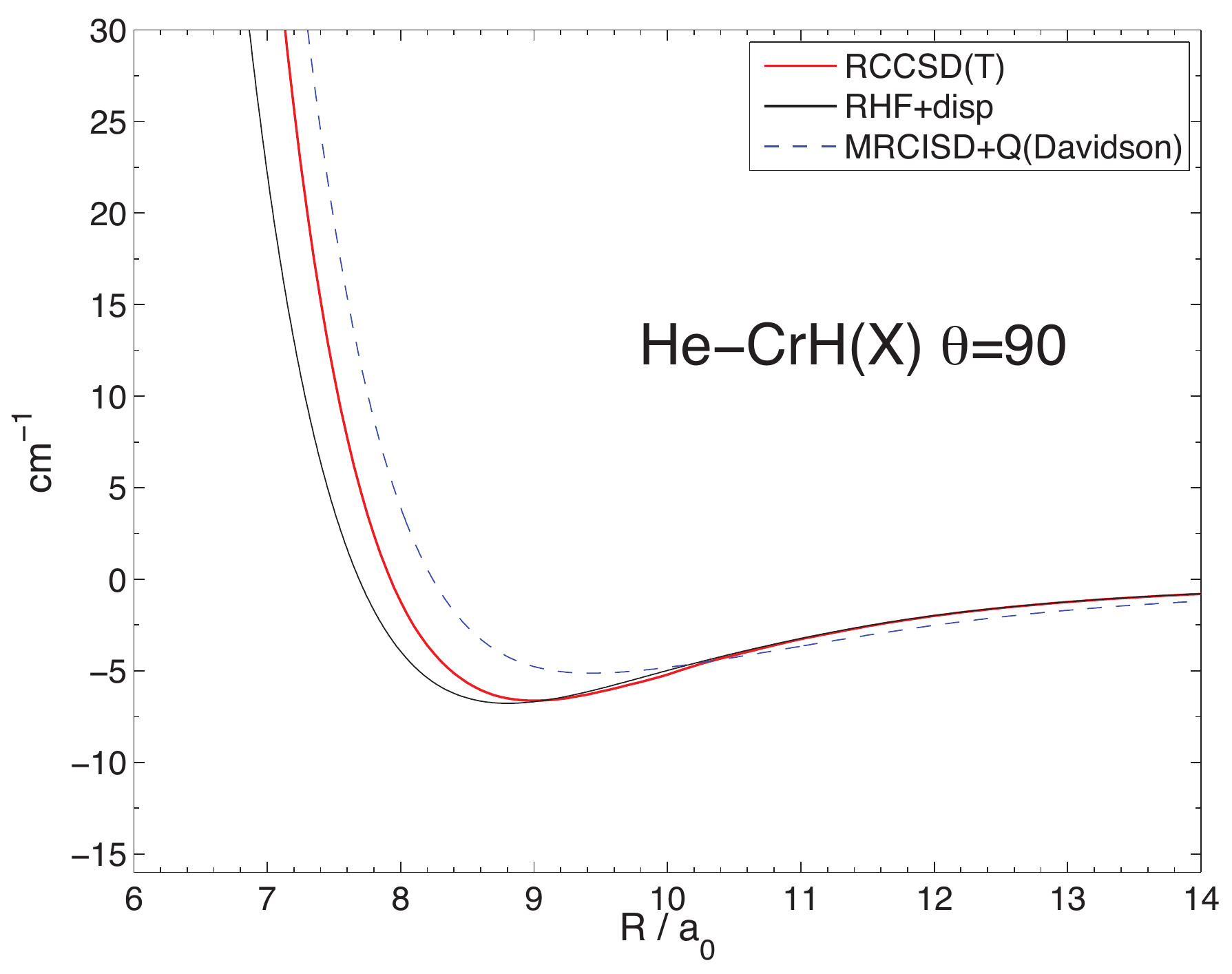}
\caption{ Comparison between RCCSD(T), {\em ic}-MRCISD+Q(Davidson) and RHF+disp potentials for the radial cut of the He-CrH(X) PES for $\theta=90$ degrees. }
\label{hecrh_sapt_90}
\end{figure}

\begin{figure}
\includegraphics[width=8.0cm]{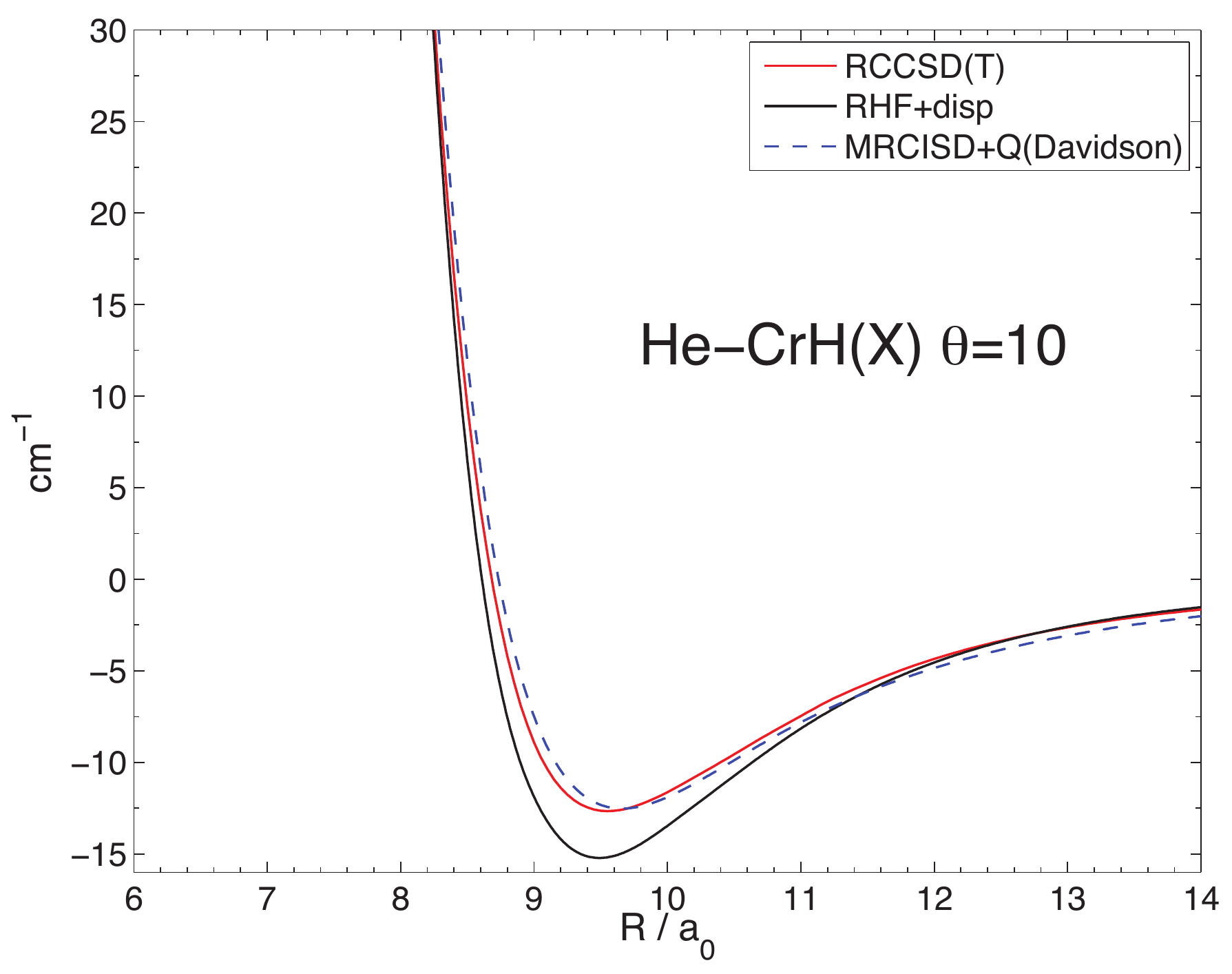}
\caption{ Comparison between RCCSD(T), {\em ic}-MRCISD+Q(Davidson) and RHF+disp potentials for the radial cut of the He-CrH(X) PES for $\theta=10$ degrees. }
\label{hecrh_sapt_10}
\end{figure}

\begin{figure}
\includegraphics[width=8.0cm]{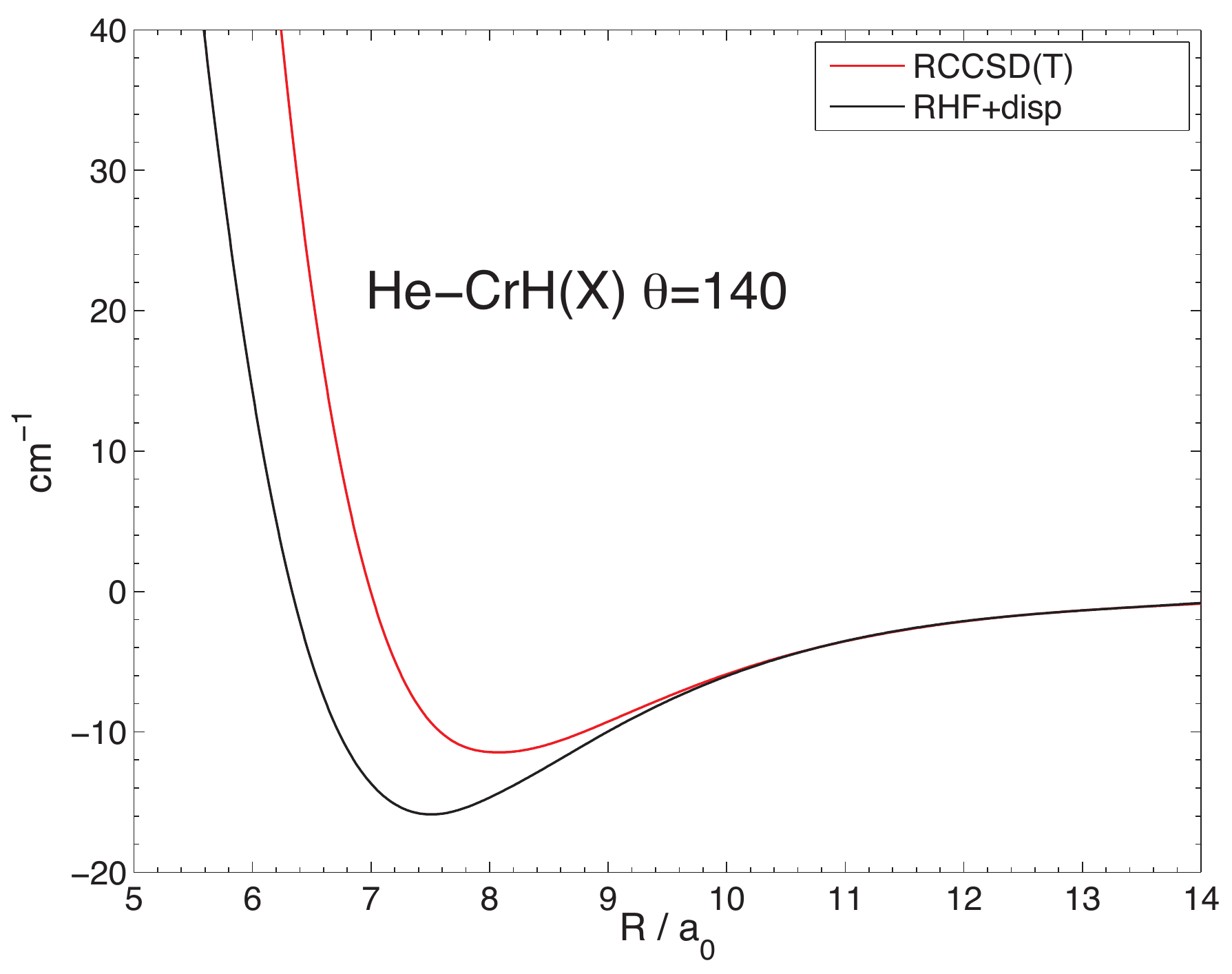}
\caption{Comparison between RCCSD(T) and RHF+disp potentials for the radial cut of the He-CrH(X) PES for $\theta=140$ degrees. }
\label{hecrh_sapt_140}
\end{figure}

In Figures~\ref{hecrh_sapt_10}, \ref{hecrh_sapt_90} and \ref{hecrh_sapt_140} we show radial cuts that compare RCCSD(T) and RHF+disp interaction energies for $\theta=10^{\circ}$, 90$^{\circ}$ and 140$^{\circ}$, respectively. The agreement is reasonable, especially good at the long range. The comparison for angles smaller than 10 and larger than 140 degrees was not possible due to the convergence problems of the RSH+disp method. This points to the methodological difficulties in obtaining correct electronic description in the vicinity of the Cr in CrH.

\begin{figure}
\includegraphics[width=8.0cm]{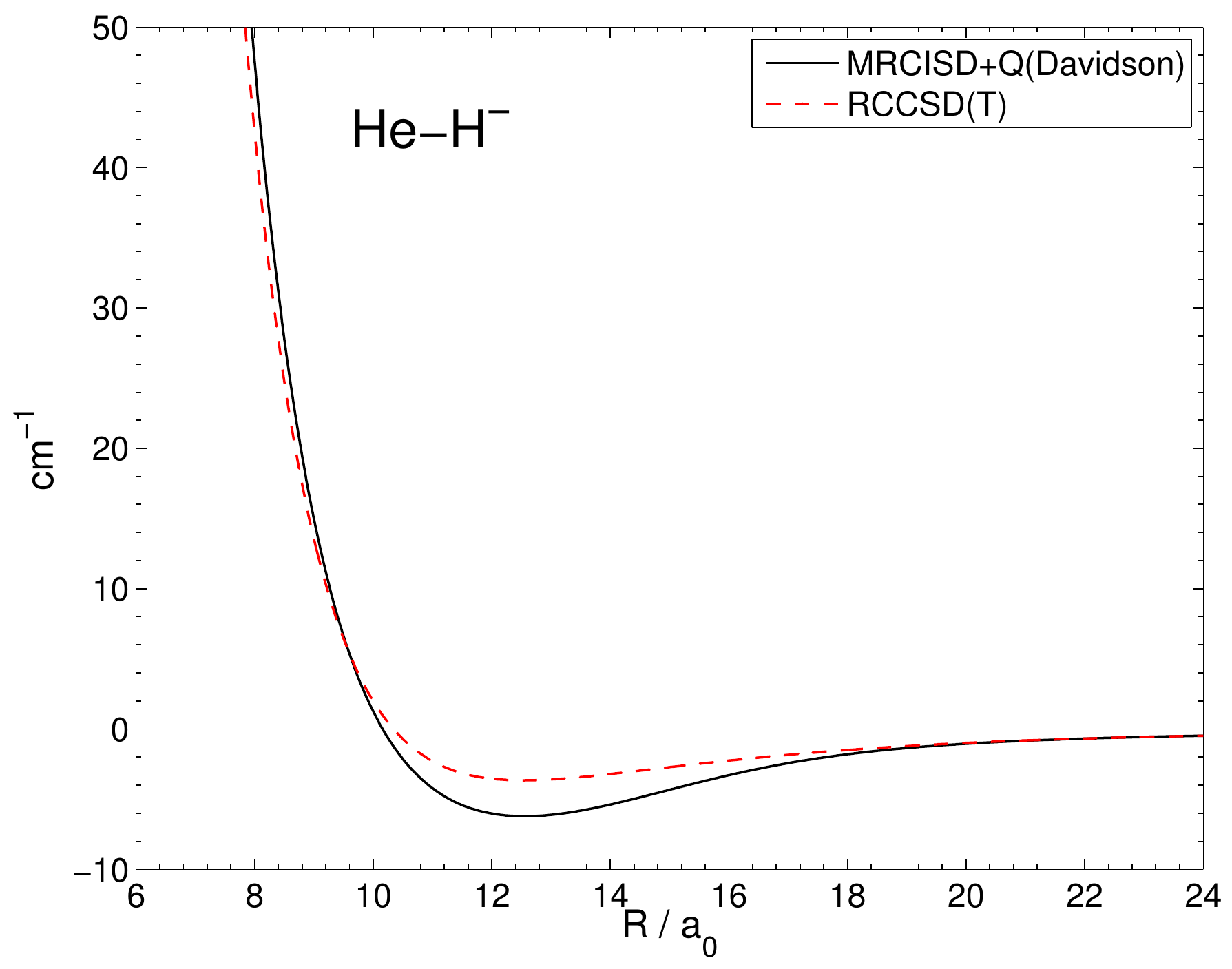}
\caption{The {\em ic}-MRCISD+Q(Davidson) (black solid) and RCCSD(T) (red dashed) potential energy curves  for the He-H$^-$($^1$S) van der Waals system. }
\label{heh_minus}
\end{figure}

The anisotropy of the He-CrH($X$) PES is unusual for a Van der Waals complex. The striking difference between the well depth of the global and local minimum amounts to more than 1143.84 cm$^{-1}$.
The reason behind such behaviour lies in the polar Cr$^+$H$^-$ charge separation, as indicated by the pronounced dipole moment of the molecule. In order to show this we calculated MCSCF/{\em ic}-MRCISD+Q(Davidson) and RCCSD(T) potential curves for He-H$^-$($^1$S) presented on Figure~\ref{heh_minus} and He-Cr$^+$($^6$S) presented on Figure~\ref{hecr_plus}. We expect those model systems to qualitatively reflect the contrasting character of the minima present in the He-CrH complex.

The minimum of the He-H$^-$ system is located at $R=13$ a$_0$ and approximately 6 cm$^{-1}$ deep in {\em ic}-MRCISD+Q calculations, 4 cm$^{-1}$ in RCCSD(T). The T$_1$ and D$_1$ diagnostics were approximately 0.01 indicating a single-reference character of this system.
The minimum for He-Cr$^+$ is present at $R=4.34$ a$_0$ with a well depth of 368.2 cm$^{-1}$ at the {\em ic}-MRCISD+Q level. Our RCCSD(T) result, $R = 4.33$ a$_0$, $D_e$ = 364.6 cm$^{-1}$, is in good agreement with the the {\em ic}-MRCI+Q results of Partridge and Bauschlicher, $R = 4.44$ a$_0$, $D_e$ = 364 cm$^{-1}$.~\cite{partridge:1994}. The T$_1$ diagnostic for this system was in the range between 0.015 to 0.019 indicating that the multireference character is slightly increased, but less than for He-CrH system.  

Both global and local minima of the He-CrH complex occur at a shorter distance and are over three times deeper than minima of their model counterparts, He-Cr$^+$ and He-H$^-$, respectively. This exhibits the substantial stabilizing role of the dispersion interaction in the He-CrH system. 
Nevertheless, the relative difference between the potentials of charge-separated systems, He-Cr$^+$ and He-H$^-$, resembles the one observed for collinear ($\theta$ = 0 and 180) cross sections of the He-CrH PES. 
  
\begin{figure}
\begin{center}
\includegraphics[scale=0.3,angle=-90]{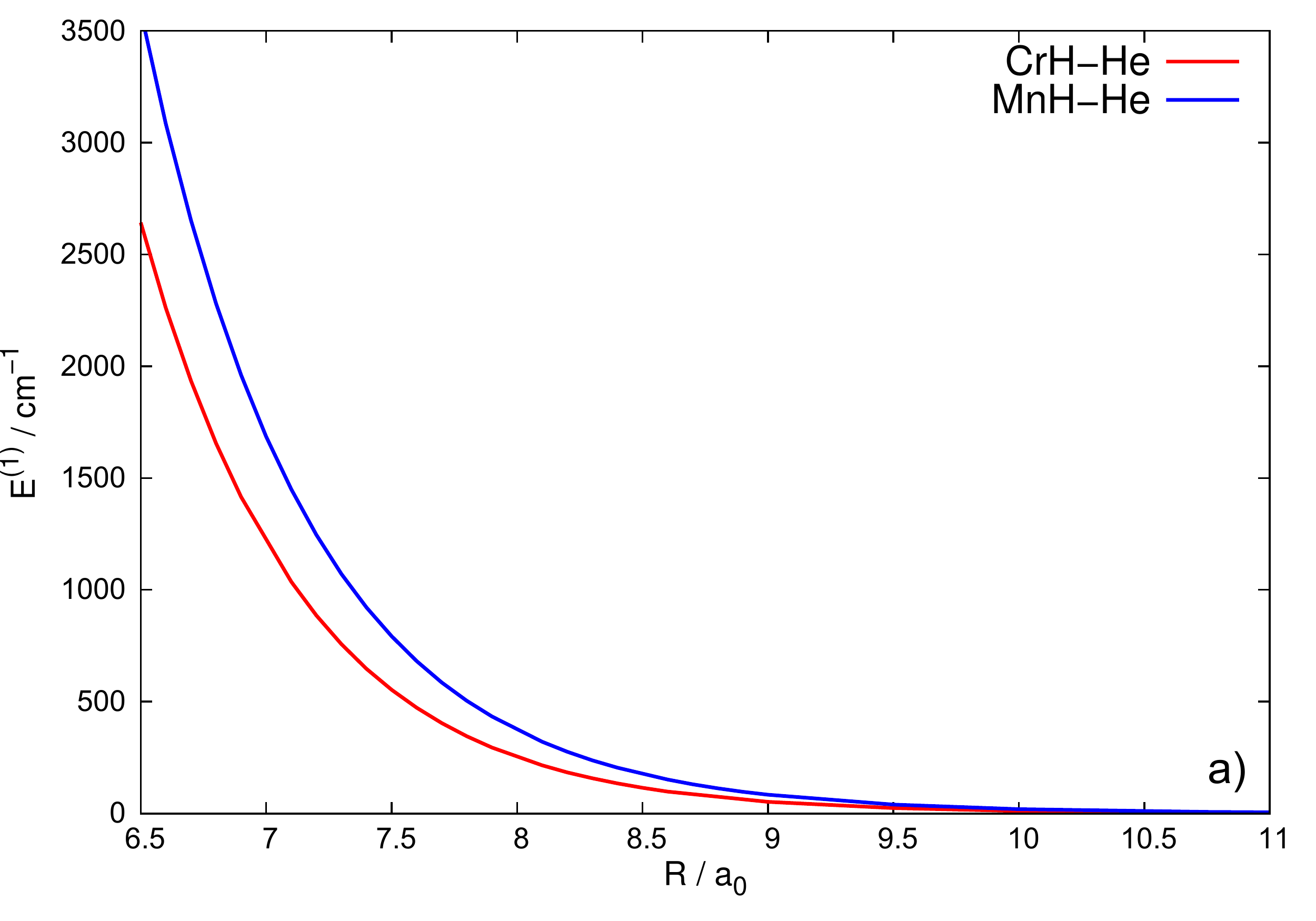}
\includegraphics[scale=0.3,angle=-90]{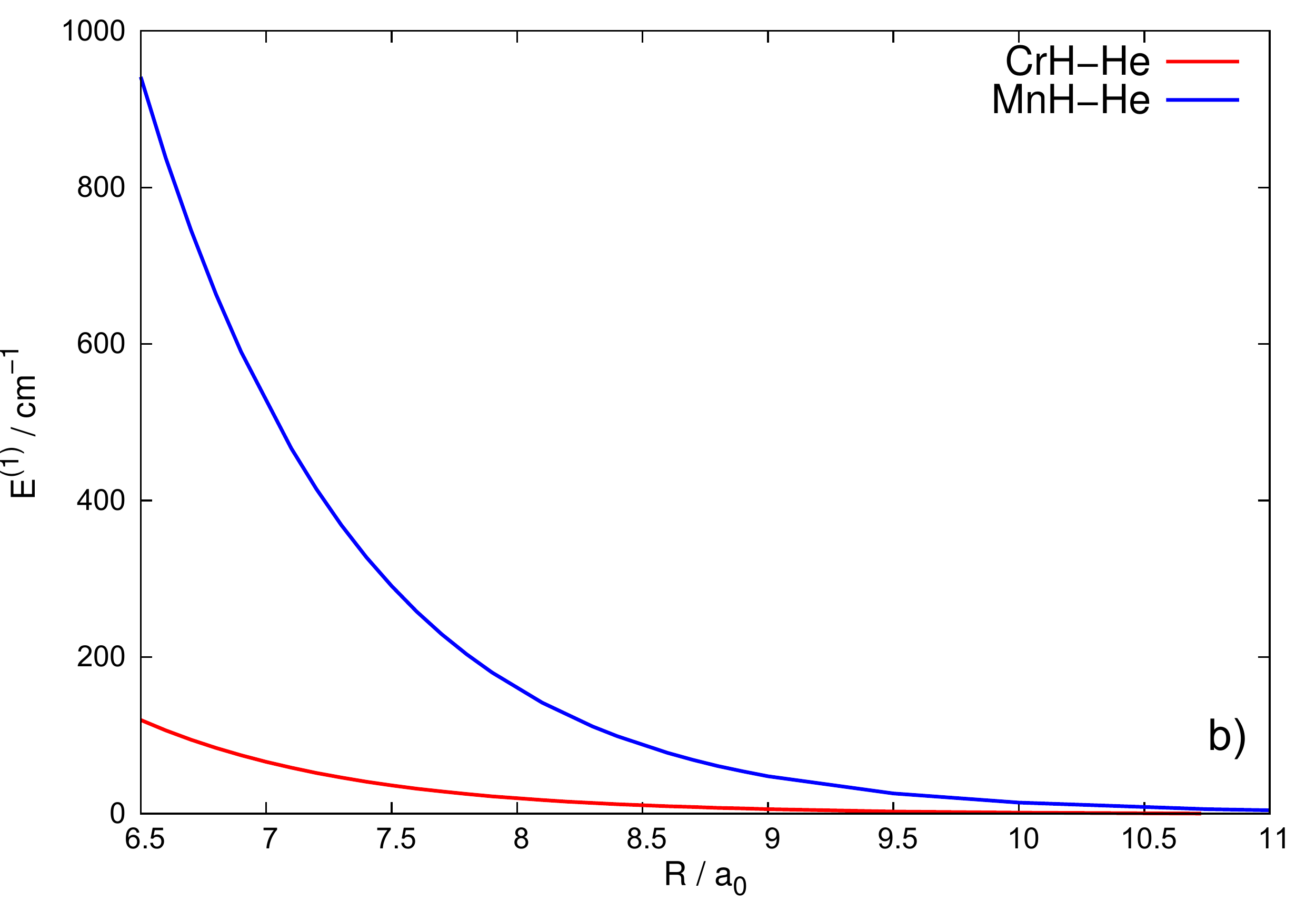}
\caption{First order SAPT energy, $E^{(1)} = E^{(1)}_{\rm elst} + E^{(1)}_{\rm exch}$ of He--CrH and He--MnH at a) $\theta = 10$ and b) $\theta = 170$.}
\label{plot:firstsapt}
\end{center}
\end{figure}

\begin{figure}
\begin{center}
\includegraphics[scale=0.5]{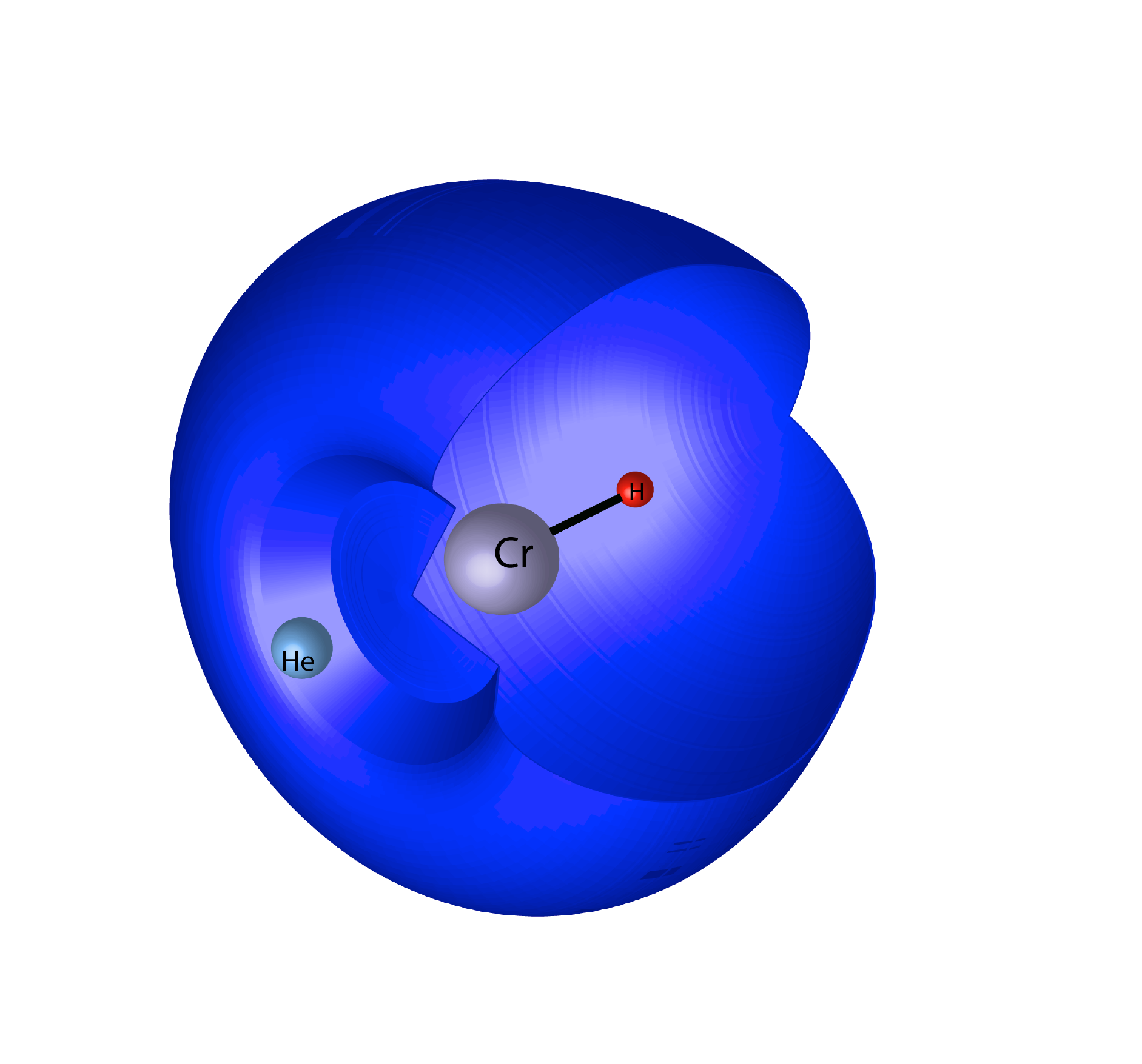}
\caption{Exchange cavity in He-CrH visualized as $E_{\rm int}$ = 0 isosurface.}
\label{plot:excav}
\end{center}
\end{figure}

Even better insight can be derived from a detailed analysis of the SAPT energy components. Such an approach reveals the presence of the so called ``exchange cavity'' -- the diminished Pauli exchange, on the one hand, and enhanced induction interaction on the chromium side of the CrH complex on the other. The presence of this phenomenon corresponds to the considerable dipole moment of the CrH molecule. It is noteworthy, that a similar effect has recently been observed for He-BeO($^1\Sigma^+$). \cite{hapka:2013} We present the ``exchange cavity'' by plotting the first order SAPT energy components, i.e. electrostatic and exchange energy, $E^{(1)}_{\rm elst}$ and $E^{(1)}_{\rm exch}$, respectively. For comparison, results for the He-MnH($^7\Sigma^+$) complex which does not exhibit such peculiar anisotropy and has much lesser dipole moment are also depicted, see Figures \ref{plot:firstsapt} and \ref{plot:excav}. \cite{halvick:2010}

\section{Bound states of the He-CrH(X$^6\Sigma^+$) van der Waals complex.}
\label{sec:bound}

We calculated bound states supported by the potential obtained in this work for total angular momentum quantum number $J=0$. The approximation of CrH as a closed shell molecule was used (no spin-splitting). We applied the collocation method on a grid composed of 30 angular points corresponding to Gauss-Legendre weights and 200 radial points spread between 2.5 and 30 a$_0$. The reduced mass of the complex $\mu=3.72129$ a.m.u. was obtained taking atomic masses of the most abundant isotopes. The rotational constant of CrH(X) was equal to $B_e=6.22$ cm$^{-1}$. 

\begin{figure}
\begin{center}
\includegraphics[width=8.0cm]{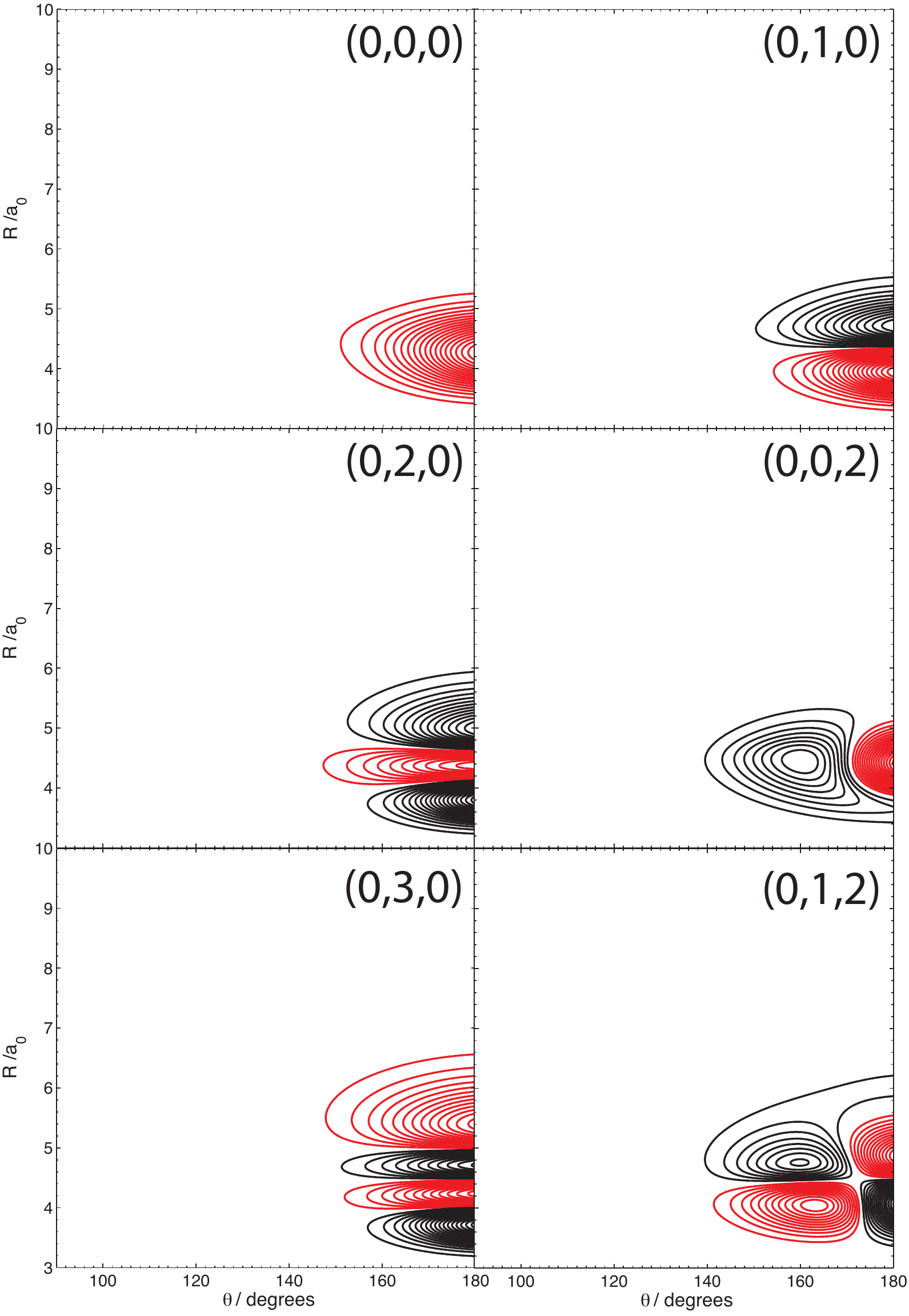}
\caption{Contour plots of first 6 selected wave functions for states labeled with $(J,\nu_s,\nu_b)$ with energies listed in Table~\ref{tab:bound}. Note that the wave functions
are localized mostly around $\theta=180^{\circ}$, therefore we plot in the $\theta$ range between 90$^{\circ}$ and 180$^{\circ}$.}
\label{fig:wavefuns}
\end{center}
\end{figure}

There are nine bound states supported by the potential for the total angular momentum $J=0$. The wave functions are mostly located in the collinear He$\cdots$Cr-H minimum and they are shown
in Figure~\ref{fig:wavefuns}. The bound state energies for $J=0$, their ro-vibrational assignments, average distance, angle values and rotational constants are shown in Table~\ref{tab:bound}.   The $D_0$ binding energy for the He-CrH system is 797 cm$^{-1}$ and the rotational constant of the complex is 0.87 cm$^{-1}$.

\begin{table} \small
\caption{\label {tab:bound} Ro-vibrational energy levels, average distance and angle and rotational constants for the He-CrH(X) complex with $J=0$. Energies and rotational constants in cm$^{-1}$}
\begin{center}
\begin{tabular}{ccccc}
\hline\hline
$(J,\nu_s,\nu_b)$ & $E_v$ & $\left<R\right>$ / a$_0$ &$\left<\theta\right>$& $B_v$\\
\hline
$(0,0,0)$  &     -796.98 &    4.338 &   165 &     0.870\\
$(0,1,0)$  &     -584.83 &    4.395 &   165 &     0.866\\
$(0,2,0)$  &     -359.99 &    4.600 &   164 &     0.808\\
$(0,0,2)$  &     -333.20 &    4.452 &   154 &     0.827\\
$(0,3,0)$  &     -185.02 &    5.020 &   164 &     0.697\\
$(0,1,2)$  &      -98.20 &    4.492 &   154 &     0.833\\
$(0,4,0)$  &      -65.32 &    5.718 &   160 &     0.553\\
$(0,5,0)$  &      -10.21 &    7.505 &   140 &     0.329\\
$(0,6,0)$  &       -1.92 &   10.922 &    90 &     0.162\\
\hline
\end{tabular}
\end{center}
\end{table}

\section*{Acknowledgments}

We acknowledge computational resources of Deepthought Supercomputer at the University of Maryland.
M.H. was supported by ``Towards Advanced Functional Materials and Novel Devices: Joint UW and WUT International PhD Programme'' Project operated within the Foundation for Polish Science MPD Programme, implemented as a part of the Innovative Economy Operational Programme (EU European Regional Development Fund).
G.C. was supported by the Polish Ministry Science of and Higher Education, Grant No. N204 248440, and by the National Science Foundation (US), Grant No. CHE-1152474. G.C. is also a beneficiary of the MISTRZ Academic Grant for Professors.

\bibliography{Bibliografia}

\end{document}